    \documentclass[lettersize,journal]{IEEEtran}
    \usepackage{booktabs}
    \usepackage{diagbox}
    \usepackage{makecell}
    \usepackage{array}
    \usepackage{graphicx,amssymb,amsmath}
    \usepackage{multicol}
    \usepackage[noadjust]{cite} 
    \usepackage{setspace}
    \usepackage{subfigure}
    \usepackage{graphicx}
    \usepackage{float}
    \usepackage {url}
    \usepackage{stfloats}
    \usepackage{amsthm,pifont}
    \usepackage{flushend}
    \usepackage{cases,subeqnarray}
    \usepackage{bm,multirow,bigstrut}
    \usepackage{amsmath, amsthm, amssymb}
    \usepackage{textcomp}
    \usepackage{latexsym,bm}
    \usepackage{booktabs}
    \usepackage{mathtools}
    \usepackage{dsfont}
    \usepackage{extarrows}
    \usepackage{epsfig}
    \usepackage{epsfig}
    \usepackage{epstopdf}
    \usepackage[table]{xcolor}
    \usepackage{array} 
    \usepackage{hyperref}
    
    \theoremstyle{plain}

    \theoremstyle{plain}

    \usepackage{amsmath}

    \IEEEoverridecommandlockouts

\begin{document}
\title{Empowering Wireless Networks with Artificial Intelligence Generated Graph}
\author{Jiacheng Wang, Yinqiu Liu, Hongyang Du, Dusit~Niyato,~\IEEEmembership{Fellow,~IEEE},
Jiawen Kang, Haibo Zhou, \\ and Dong In Kim,~\IEEEmembership{Fellow,~IEEE}

\thanks{J.~Wang, Y.~Liu, H.~Du and D.~Niyato are with the College of Computing and Data Science, Nanyang Technological University, Singapore (e-mail: jiacheng.wang@ntu.edu.sg, yinqiu001@e.ntu.edu.sg, hongyang001@e.ntu.edu.sg, dniyato@ntu.edu.sg).}
  
 \thanks{J. Kang is with the School of Automation, Guangdong University of Technology, China (e-mail: kavinkang@gdut.edu.cn).}
 

 \thanks{H. Zhou is with the School of Electronic Science and Engineering, Nanjing University, Nanjing, China (haibozhou@nju.edu.cn).}

 \thanks{D.~I.~Kim is with the Department of Electrical and Computer Engineering, Sungkyunkwan University, Suwon 16419, South Korea (email:dikim@skku.ac.kr).}




}

\maketitle
\begin{abstract}
In wireless communications, transforming network into graphs and processing them using deep learning models, such as Graph Neural Networks (GNNs), is one of the mainstream network optimization approaches. While effective, the generative AI (GAI) shows stronger capabilities in graph analysis, processing, and generation, than conventional methods such as GNN, offering a broader exploration space for graph-based network optimization. Therefore, this article proposes to use GAI-based graph generation to support wireless networks. Specifically, we first explore applications of graphs in wireless networks. Then, we introduce and analyse common GAI models from the perspective of graph generation. On this basis, we propose a framework that incorporates the conditional diffusion model and an evaluation network, which can be trained with reward functions and conditions customized by network designers and users. Once trained, the proposed framework can create graphs based on new conditions, helping to tackle problems specified by the user in wireless networks. Finally, using the link selection in an integrated sensing and communication (ISAC) as an example, the effectiveness of the proposed framework is validated.
\end{abstract}

\IEEEpeerreviewmaketitle
\section{Introduction}

Graphs play an essential role in the interconnected world, serving as fundamental structures for representing and analyzing complex systems across various fields. For example, in biology, graphs illustrate complex interactions within an organism, from the cellular level to the entire ecosystems, enhancing our understanding of life's processes~\cite{guo2022systematic}. In logistics and transportation, they model networks such as roads and flight paths to improve the efficiency of transporting goods and people. The widespread applications of graphs demonstrate their significance as essential tools for addressing real-world challenges across various domains.

In the context of wireless communications networks, the importance of graphs is especially highlighted due to the networks' dynamic and distributed nature. Within these networks, the nodes can represent devices such as routers, base stations, and mobile users, while the edges illustrate the potential communication links among devices. Therefore, graphs not only visualize the network topology but also serve as analytical tools for performance optimization, fault detection and recovery, and scalability analysis~\cite{shen2022graph}. Regarding these topics, extensive and in-depth research has been conducted, primarily utilizing models such as graph neural networks (GNNs) and deep learning. However, an emergence of technologies such as integrated sensing and communications (ISAC) and semantic communications, has introduced a range of new challenges that traditional models may struggle to address. Generative artificial intelligence (GAI) can learn the distribution features of given graphs and then generate new graphs based on the specific conditions, offering novel approaches to solve these emerging problems.

In wireless networks, producing graphs using GAI involves several key steps. First, it requires the collection and preprocessing of training data, such as the topologies of existing networks, traffic patterns, and node characteristics. Then, this data is used to train the GAI models, such as generative adversarial networks (GANs) or variational auto-encoders (VAEs), which learn the underlying distribution of the network graphs. During training, the model parameters are adjusted to accurately capture the complex relationships and dynamics within the wireless network. After training, these models can then generate new graphs, which play several pivotal roles in the design and operation of wireless networks as shown below~\cite{he2021overview}.
\begin{itemize}
    \item \textbf{Simulating Complex Scenarios}. Generated graphs can simulate various network conditions, such as high traffic loads, node failures or attacks. These simulations are invaluable for network stress testing and preparation for real-world challenges. 

    \item \textbf{Network Planning and Design}. By generating a wide range of potential topologies, network designers can explore various configurations to find the most efficient layout. This exploration might include optimizing the placement of base stations, designing fault-tolerant architectures, and so forth.

    \item \textbf{Dataset Augmentation}. Generated graphs can serve as training and testing datasets for machine learning models. These models can predict network failures, detect unauthorized access, or dynamically adjust network parameters for optimal performance.

    \item \textbf{Exploring What-if Scenarios}. Generated graphs allow network to explore unknown scenarios, such as impacts of adding or removing nodes, changing traffic patterns, or implementing new network protocols. This supports optimal decision-making and strategic planning in complex network environments.
 
\end{itemize}

As can be seen, graph generation plays a multifaceted role in wireless communications networks. It can enhance performance of the network itself, such as optimizing network topologies for better signal coverage and managing interference among nodes to improve communication quality. Additionally, graph generation can provide support for network stress tests and new structure exploration. These contributions are vital for creating more reliable and cost-efficient network architectures.

Considering the pivotal role of graphs in wireless networks and the new opportunities brought by GAI, this paper explores applications of GAI-based graph generation in wireless networks. First, we present a comprehensive discussion on the applications of graph in various domains, such as drug molecule generation, routing optimization, and resource allocation. Then, we introduce and analyze commonly-used GAI models for graph generation. Finally, we present a diffusion based graph generation framework and illustrate how this framework can support the selection of sensing and communication links in ISAC systems via a case study. The main contributions of this paper are summarized as follows.

\begin{itemize}
\item We survey and analyse the applications of graphs from various perspectives, including routing optimization and resource allocation, which provide background and highlight benefits of GAI-based graph generation in wireless networking.

\item We introduce the major GAI models from the perspective of graph generation, including their principles, strengths, and weaknesses. These fundamentals provide a comprehensive review of using different GAI techniques to achieve different objectives of graph generation.

\item We propose a diffusion-based conditional graph generation framework and validate its effectiveness through the case study of the link selection in the ISAC system.
\end{itemize}

\section{Applications of Graph Generation}
This section discusses the applications of graph in the fields of molecular design and wireless communication networks.
\subsection{Applications in Molecular Design}
\subsubsection{Drug Molecule Generation}
In drug generation, GAI models showcase advanced capabilities in designing and optimizing molecular structures. For instance, GraphAF~\cite{shi2020graphaf} employs an auto-regressive (AR) model to build molecular graphs, ensuring structural validity through reinforcement learning. It excels in creating molecules with enhanced pharmacokinetic properties, such as boosting drug absorption and distribution, showcasing its utility in creating more effective and efficient drug candidates. In addition to ARs, VAEs are also widely adopted. For example, the junction tree VAE (JT-VAE)~\cite{godinez2022design} realizes the structure generation by breaking down molecules into a junction tree, where each node represents a molecular substructure or motif, while the edge depicts the connection among them, as shown in Fig.~\ref{APP}. JT-VAE can generate molecules with high solubility and metabolic stability by progressively adding effective chemical substructures, which guarantees both the structural soundness and chemical accuracy of the molecules. GAI models present a robust method for creating new and inventive molecules, showing the potential to revolutionize the fields of pharmaceutics.

\subsubsection{Protein Structure Design} 
The protein can be encoded as a graph, where nodes represent amino acids and edges depict interactions among them. Therefore, GAI can be trained based on the protein graphs to produce the novel protein structure. For instance, the authors in~\cite{wu2024protein} introduced a diffusion model based method, aiming to generate protein backbone structures, as shown in Fig.~\ref{APP}. They defined the structure of protein backbones through a sequence of angles that detail the relative orientations of backbone atoms. The generation process starts with a random and unfolded state, and then progressively removes noise to reach a stable and folded conformation. The key advantage of this method is the use of shift and rotational invariant representations, which reduces the need for complex equivariant networks. Utilizing the transformer as the denoising diffusion probabilistic model, they demonstrated that the model can unconditionally generate highly realistic protein structures with complexity and structural patterns similar to those of naturally-occurring proteins.
\subsection{Applications in Wireless Communication Networks}
\subsubsection{Routing Optimization} 
The nodes and dynamic interplay of data flow within the network can be mapped as graphs. By analyzing these graphs, GNNs can predict potential congestion and optimize the allocation of data pathways. Such predictive capability is rooted in the accurate depiction of the network by the graph and GNNs' ability to learn from the network's historical data patterns. For example, by taking advantage of the extensive data available from the controller of a software-defined network (SDN), the authors in~\cite{swaminathan2021graphnet} introduced a routing algorithm powered by GNN. This algorithm aims to predict the most efficient routing path that minimizes the average delay between the source and destination nodes within SDN. Additionally, a deep reinforcement learning framework was developed to train the GNN using prioritized experience replay from the experiences learned by the controllers. The algorithm was evaluated based on various topologies in terms of packets successfully routed and average packet delay time, validating its robustness to the changes in network structure and the varying hyperparameters. 

\subsubsection{Resource Allocation}
\begin{figure*}[t]
	\centering
	\includegraphics[width=0.95\textwidth]{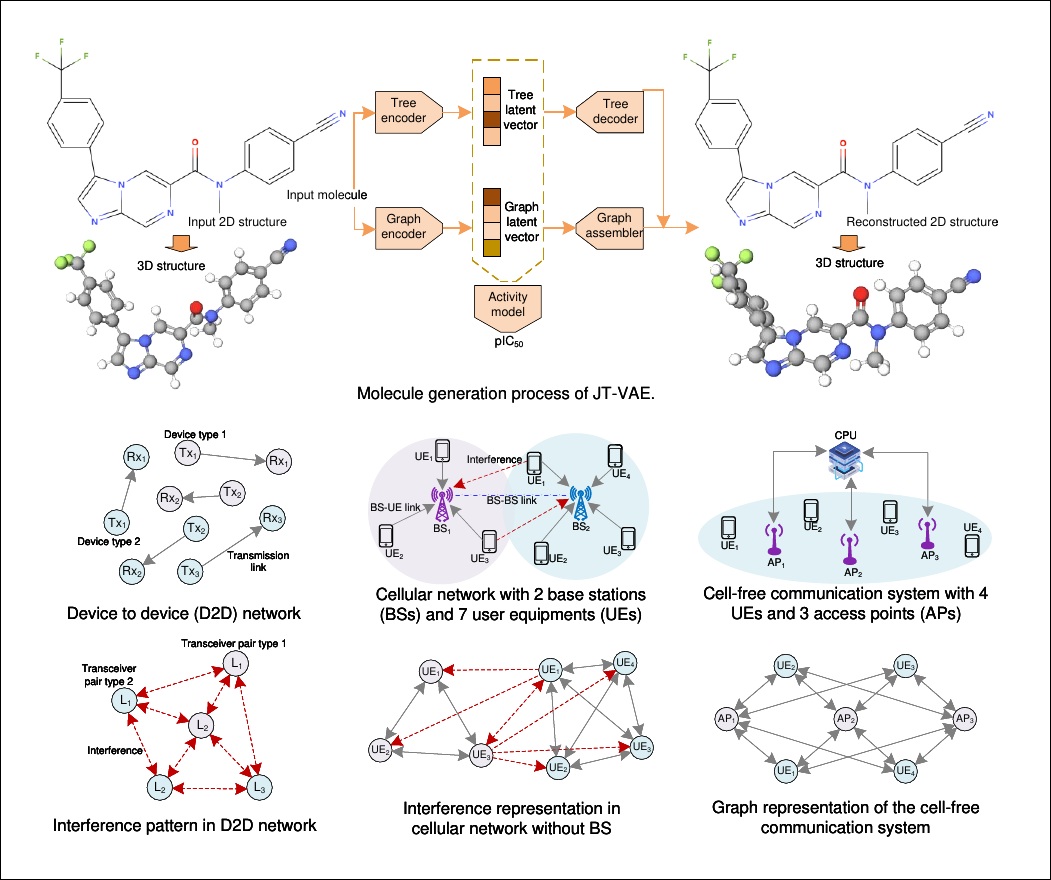}%
	\caption{The applications of graph in molecule generation, D2D communications, cellular networks, and cell-free communication systems. In JT-VAE, An input molecule undergoes encoding via a tree and graph encoder, which generates corresponding tree and graph latent vectors. These vectors are then utilized by a tree decoder and a graph assembler for the purpose of reconstructing the molecule. Additionally, the tree and graph latent vectors are combined and used as input for a model that predicts the pIC50 value of the input molecule. In D2D communication and cellular networks, graphs offer a structured way to represent the transmission links between devices and base stations (BSs), enhancing the management of link interference. In cell-free communication systems, graphs simplify the connections between user equipment (UEs) and access points (APs). This simplification aids in optimizing cooperative transmission and reception, improving power control, and facilitating efficient resource sharing across the network.}
	\label{APP}
\end{figure*}

In wireless communications, networks can be represented as graphs, where nodes represent entities such as base stations, mobile devices, or sensors, and edges signify the connections or communication paths between these entities. GNNs can process these graphs by aggregating and updating node information through the network layers, thereby enabling efficient resource allocation. For instance, in~\cite{shen2020graph}, the authors formulated the radio resource management as graph optimization problems. Then, they proposed a novel message passing GNNs (MPGNNs), which are notable for their ability to handle permutations, scale to larger problems, and deliver high computational speed. They proved the equivalence between MPGNNs and a family of distributed optimization algorithms, which were then used to analyze the performance and generalization of MPGNN-based methods. Using power control and beamforming as case studies, they showed that the MPGNNs outperform the traditional optimization approaches, all without requiring domain-specific knowledge. Besides routing and resource allocation, the graph can also be used in device to device (D2D) communication, cellular network, and cell-free communications, which are presented in Fig.~\ref{APP}.

\subsection{Lessons Learned}
From the applications discussed above, we can summarize several key points.

\begin{itemize}
\item By abstracting data into nodes (entities) and edges (relationships), graphs provide a foundational method for encoding complex entities and their interactions across different domains. This is crucial for transforming intricate systems and issues into manageable ones. 

\item The graphs can be customized to address specific challenges inherent to diverse fields, such as biology, chemistry, and wireless communications. This flexibility enables researchers to enhance a system by optimizing the graph according to the unique requirements, while ensuring that the models are relevant and effective. 

\item Unlike linear models, graph-based approaches can easily represent and analyze the interplay between different entities and their connections. This ability to navigate complex relational spaces facilitates the discovery of innovative structures and solutions, facilitating relevant research and application development.
\end{itemize}

From these applications and key points, it can be observed that GAI demonstrates strong generalization in graph generation, which has been widely applied in areas such as molecular structure design. However, in wireless communications, existing works mainly rely on conventional GNNs. While offering impressive computing efficiency, they still face some limitations. For instance, the generalization of GNNs in radio resource management relies on maintaining the permutation equivariance property in graph representations. They can be applied to varied problem scales when this property is preserved, but they may struggle if the network structure deviates from the training scenarios. This is particularly evident when models trained on the dataset with a fixed number of users are applied to real-world situations where the number of users varies. Therefore, it is imperative to introduce GAI for graph generation to further enhance the capabilities in various aspects of wireless communications.

\section{Generative AI Models for Graph Generation}
This section introduces the commonly used GAI models from the perspective of graph generation, including basic principles, advantages and disadvantages, etc.

\subsection{Auto-Regressive Model}
AR models utilize a sequential generation strategy, in which each step predicts the next node or edge based on the previously generated graph. The commonly used objective function is the maximum-likelihood estimation (MLE), steering the model to create sequences that mirror those in the training set. It relies on the chain rule of probability to factorize a joint distribution over the random variables, necessitating a pre-specified ordering of nodes within the graph to guide the generation process. For example, GraphRNN~\cite{you2018graphrnn} utilizes Recurrent Neural Networks (RNNs) to learn and generate the distribution of various graph representations, adhering to node orderings. These RNNs sequentially generate nodes and edges, capturing the intricate structural dependencies inherent to the graph. This sequential generation is suited for the network topology creation that can potentially optimize parameters such as connectivity, signal coverage, and network resilience against failures. Moreover, the ability to model and generate network topologies enhances wireless network simulation and planning, facilitating the creation of networks that cater to the rising demands for bandwidth and connectivity in various environments. While impressive, the reliance on a sequential generation process introduces challenges, particularly in designing permutation-invariant graph distributions, as they inherently treat the graph as a sequence.

\subsection{Variational Auto-encoders}
VAEs employ a dual-component architecture, where an encoder compresses input data into a compact latent space and a decoder reconstructs data from this latent representation back to its original form. It estimates the distributions of graphs by maximizing the evidence lower bound (ELBO), i.e. maximizing the difference between the reconstruction loss and the disentanglement enhancement term. VAEs mainly generate graph through two methods, including one-shot and sequential method. The one-shot method predicts edge probabilities by analyzing node embedding interrelations. GraphVAE~\cite{simonovsky2018graphvae}, for example, employs a decoder to yield a probabilistic, fully-connected graph in a single action. It utilizes approximate graph matching to determine the optimal graph structure, hence ensuring that only particular node and edge combinations are considered valid. Unlike one-shot method, sequential generation constructs graphs node by node, making decisions at each step about connectivity and edge attributes to progressively build the graph. Through these methods, VAEs enable more accurate network simulations, leading to improvements in network capacity, resilience against fluctuations in demand, and overall performance stability. While promising, they face scalability challenges in generating large-scale graphs, complicating their use in applications demanding extensive graph structures.

\subsection{Normalizing Flows}
Normalizing flows (NFs) work through invertible transformations. Specifically, they first transform the graph's probability distribution into a more complex one using a series of invertible transformations via an encoder. Following this, a decoder reverses these transformations to generate the graphs. Typically, these models are trained by minimizing the negative log-likelihood over the training data, ensuring that the flow from simple to complex and back to simple is smooth and efficient. For instance, GraphNVP~\cite{madhawa2019graphnvp} generates graph via a two-step process, where it first generates a graph structure, and then assigns node features based on the generated structure. This allows for the generation of molecular graphs by modeling the distribution of graph embeddings obtained through the encoder. Besides, the invertible nature permits a detailed density estimation, which is particularly useful for modeling the distribution of network traffic, identifying potential bottlenecks, and optimizing the flow of data through the network. However, the necessity of invertibility and differentiability constrains the design of these models, potentially limiting their ability to accurately capture some complex data distributions. 

\subsection{Generative Adversarial Networks}
GANs are implicit generative models that learn to sample real graphs. GANs include a generator for generating graphs and a discriminator for distinguishing between generated and real graphs. During training, these two modules compete with each other until an equilibrium is reached, at which point the discriminator cannot distinguish between real and generated graphs. For instance, the authors in~\cite{de2019molgan} use GANs for generating small molecular graphs, with the generator creating probabilistic graphs in one step and the discriminator identifying whether the molecule graph is generated or a real one. It also employs a reward network for reinforcement learning, optimizing for desired chemical properties. For wireless networks, the dynamic competitive process allows GANs to model complex network behaviors and topologies effectively, making them useful for optimizing network configurations. Besides, GANs can simulate various conditions, from typical operational states to extreme stress tests, enhancing the robustness and efficiency of network deployment. However, this adversarial training can be unstable and requires careful tuning of hyperparameters and architecture to reach the equilibrium, which might be hard to achieve.

\subsection{Denoising Diffusion Model}
Unlike previous models, diffusion models operate by gradually adding and then reversing noise on a graph distribution through a series of steps. This approach allows for the generation of complex graph structures by iteratively denoising. Diffusion models have been used for creating intricate molecular graphs. These models initiate with a simple distribution and incrementally introduce noise. Subsequently, they employ a reverse process to refine the noisy distribution into a graph with desired molecular characteristics~\cite{zhang2023survey}. By applying noise in a controlled manner and then methodically reversing it, diffusion models can mimic the effects of environmental factors, hardware failures, or cyber-attacks on the network infrastructure. This capability allows for the testing of network resilience and the effectiveness of recovery protocols in a simulated environment, providing valuable feedback for improving network robustness and fault tolerance. However, the iterative nature of the model demands significant computing resources, and the balance of noise addition and removal requires careful calibration to prevent the loss of essential graph features.

\begin{figure*}[t]
	\centering
	\includegraphics[width=0.95\textwidth]{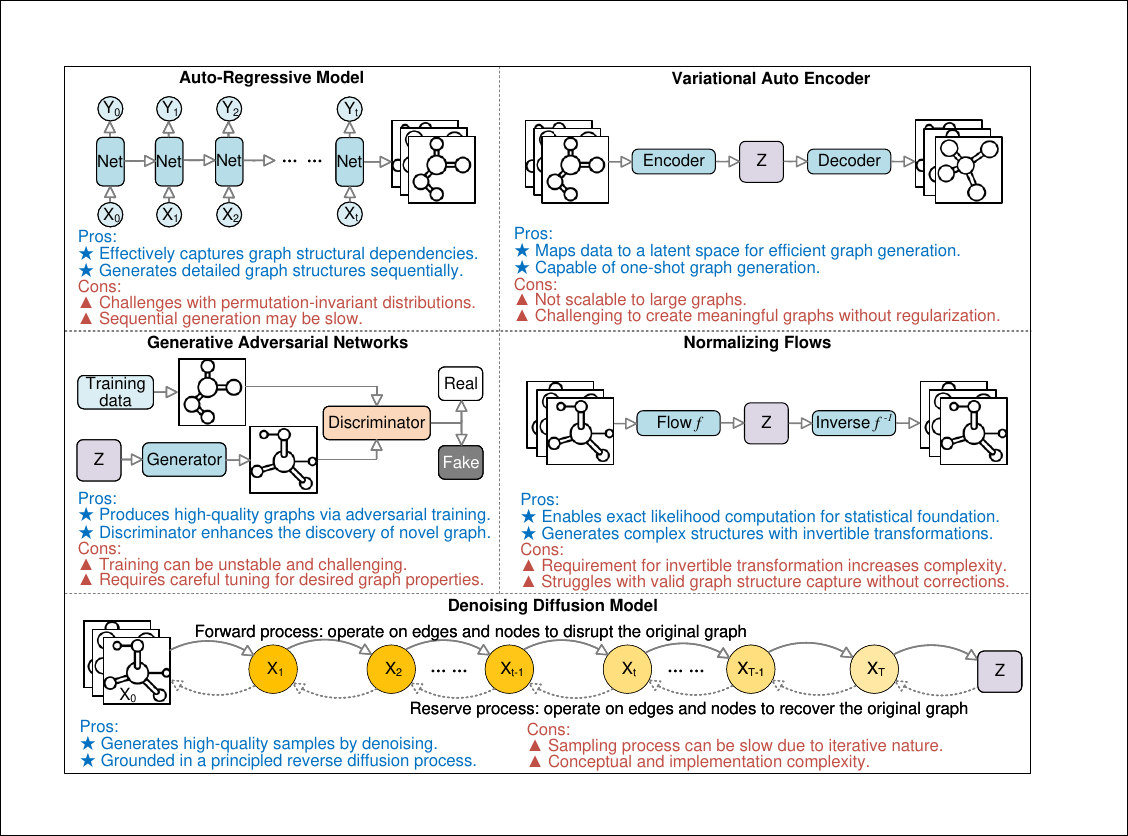}%
	\caption{The summary of GAI models from the perspective of graph generation.}
	\label{GAIM}
\end{figure*}

Figure~\ref{GAIM} provides a brief summary of the models discussed above. From the adversarial training of GANs to the meticulous step-by-step improvement of denoising diffusion models, analyzing these GAI models uncovers various methods for graph generation. Each model carries its unique advantages and challenges, but together, they significantly enhance the capability to construct and refine graphs for complex wireless communication network architectures.
\section{Graph Generation for Wireless Network}

\subsection{Motivations and Challenges}
GAI models offer a suite of tools suited to generate graphs for wireless networks. For example, AR models hold the potential to predict and adapt to time-variant changes in network traffic and node availability, ensuring stable connectivity even under variable conditions. VAEs and denoising diffusion models could facilitate the design of resilient network topologies capable of maintaining optimal performance amidst fluctuating signal strengths and environmental interference. Therefore, integrating GAI technology to generate and optimize graphs for wireless networks combines the advantages of graph-based approaches with GAI's advanced generation capabilities, providing solid support for wireless communication frameworks. However, there are still various challenges in applying GAI to generate graphs for wireless networks.
\begin{itemize}
\item \textit{Dataset Availability and Quality}: The effectiveness of GAI models heavily depends on the availability of high-quality, representative data. In wireless networks, collecting comprehensive datasets that accurately reflect the network's multifaceted nature is challenging, impacting the models' training and performance.

\item \textit{Computing Resources and Efficiency}: GAI models, particularly those involving iterative processes or complex simulations, require significant computing resources. The real-time or near-real-time analysis in wireless networks requires efficient computational strategies without compromising model accuracy or depth, which further exacerbates this challenge.

\item \textit{Adaptation and Stability}: Wireless networks necessitate maintaining a high level of robustness and reliability in dynamic environments. This requires GAI models to be adaptable, enabling stable solution generation in response to changing conditions.
\end{itemize}

The exploration of GAI models based graph generation for wireless networks highlights both the exciting potential and the intricate challenges. From addressing data quality issues to ensuring computing efficiency and model adaptability, the path to fully leveraging GAI in wireless networks comes with obstacles that needs innovative solutions.

\subsection{The Proposed Framework}
Given the advantages of diffusion models in constraint handling, training stability, and scalability~\cite{liu2024graph}, this article introduces a framework based on conditional diffusion models for generating graphs to support wireless networks. The framework comprises a denoising network and an evaluation network. During training, the input to the denoising network includes random noise, denoising time steps, and conditions, where the conditions are represented as vectors or matrices, determined by the specific problem. Unlike the continuous diffusion models used for image generation, the denoising process in this framework is discrete. Therefore, before feeding the random noise to the denoising network, the one-hot encoding is used to convert the noise into graphs organized by nodes and edges, ensuring a discrete denoising process. The denoising process starts with the noise graph and, through multiple denoising steps, such as adding or removing nodes or edges between nodes, generates the required graph. The denoising is conceptualized as a Markov decision process, with each state representing a graph at a specific diffusion time step, and actions corresponding to transitions between states. The policy, in this context, is represented by the learned conditional probabilities that dictate how denoising should be performed at each step.
\begin{figure*}[t]
	\centering
\includegraphics[width=1\textwidth]{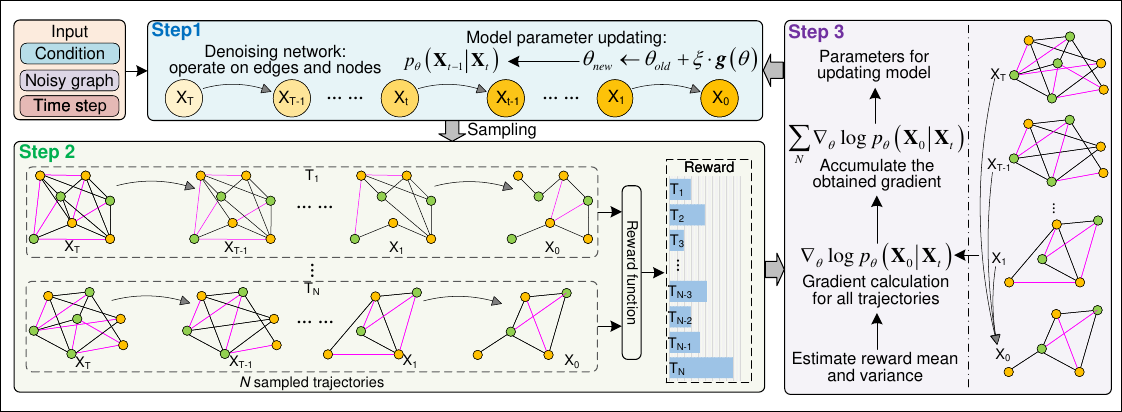}%
	\caption{The structure of the proposed framework. Our framework comprises two core modules, i.e., the denoising network and evaluation module. During training, the first step is to input the noisy image, the denoising time step, and conditions, based on which the denoising network generates the graphs. In Step 2, the defined reward function is used to evaluate the generated graphs. Finally, in Step 3, the cumulative gradient is calculated based on the reward and the result is fed back to the denoising network for optimizing its parameters. Once trained, the denoising network is capable of generating graphs based on newly input conditions.}
	\label{FW2}
\end{figure*}

The output of the denoising network is a graph with nodes and edges possessing categorical attributes, where nodes and edges represent different terminals and wireless links in a wireless network, respectively. Based on this, a function is defined to calculate the reward for each generated graph. Similar to the conditions, the reward function also needs to be defined according to the specific problem. The calculated reward is then fed into the evaluation network, which computes the cumulative gradient of the generation trajectories of sampled graphs. Essentially, the calculated gradient represents the direction and magnitude by which the parameters of the denoising network should be adjusted to increase the likelihood of generating the desired graphs. Therefore, this gradient is fed back to the denoising network for the update of network parameters. After completing training, the denoising network can generate graphs based on the input conditions, thereby addressing the specified problems in wireless networks. The overall structure of the proposed framework is depicted in Fig.~\ref{FW2}. Compared with methods that require training datasets, our framework, through the reward function and evaluation network, facilitates training in a manner more amenable to problems in wireless networks, where collecting sufficient training data is challenging. Furthermore, unlike mere imitation generation, our framework also incorporates generation conditions as the input, endowing it with stronger generalization capabilities.

\section{Graph Generation for Link Selection in ISAC Network}

\subsection{Case Study}
\subsubsection{Experimental Configurations} 
ISAC systems represent a major development in wireless communications by combining sensing and data transmission into a single framework. The penetration of wireless communication devices forms a large ISAC network, but activating all devices for sensing and communication consumes unnecessary resources including transmitting power and bandwidth. Hence, we apply the proposed framework to generate graphs, guiding the ISAC network to activate devices and links for sensing and communication under different conditions, ensuring effective and economical operation. Consider nine wireless devices forming an ISAC network under line-of-sight conditions\footnote{For ease of presentation, this article considers only a scenario involving nine devices. Nonetheless, the proposed framework can be applied to a large scale network.}. Within the coverage area, there is a dynamic target to be sensed. Treating ISAC devices and communication links as nodes and edges of the graph, respectively, and using the target's location as the generating condition, the proposed framework can generate the graph that guides an appropriate activation strategy of devices and links in the ISAC network for sensing and communication.

During the training, a generation condition is set as the location of the target, aiming to minimize node and link activation while ensuring that the target is within or near the first Fresnel zone of the activated links. Therefore, the reward function is defined as the difference between the gain associated with the target's distance to the first Fresnel zone formed by the activated links and the cost of activating the links. Here, the Fresnel zone is a series of confocal prolate ellipsoidal regions of space between and around the transmitter and receiver~\cite{wang2024unified}. When the target is in the first Fresnel zone, i.e., the innermost elliptical area, the sensing performance is better. 

We set the signal frequency at 2.4 GHz, allowing us to calculate the area of the first Fresnel zone of each activated link. Such a zone is an ellipse, with the major axis being the distance between nodes and the minor axis determined by the first Fresnel radius, which is computed based on the signal wavelength and the distance between the signal transmitter and receiver. The cost of activating a link is set as a constant, as the signal transmitting power of different ISAC devices is fixed\footnote{\url{https://lancelot1998.github.io/Graph\_Networking}}. After training, the denoising network uses the target location as generation condition to generate graphs, which guide the ISAC network to selectively engage nodes and links for target sensing, ensuring resource-efficient operation.

\subsubsection{Performance Analysis} 
\begin{figure}[tbp!]
  \centering
  \includegraphics[height=6.5cm]{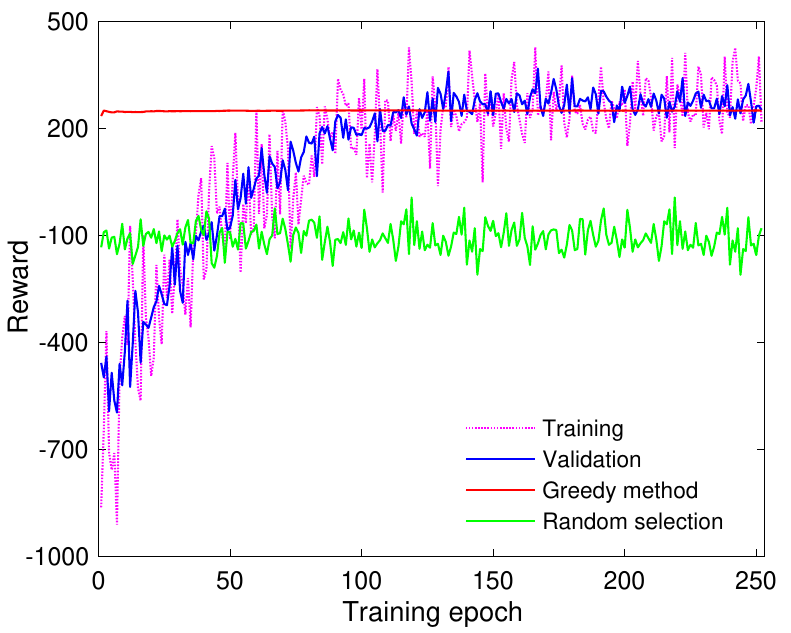} 
  \caption{The training curve of the proposed method and the comparison with other methods.} 
  \label{TR}
\end{figure}

 In Fig.~\ref{TR}, we present the training and validation curves of the proposed method and compare them to those of the random selection and greedy methods. Here, the greedy method selects the four nodes closest to the target, forming links for target sensing. The experimental results indicate that the proposed method converges after 150 epochs, achieving an average reward of approximately 300. This outperforms the random selection and greedy methods, which yield average rewards of about -100 and 250, respectively. These results show that our framework can successfully optimize the denoising network's parameters by using the gradient feedback from the evaluation network. This allows the network to learn the rules for generating graphs, which are set by the user-defined reward function, under the specified conditions. Furthermore, the validation curve demonstrates that the model can use the learned generation rules to produce graphs that maximize the reward value based on newly input conditions, thereby providing support for activating links in the ISAC network.

\begin{figure*}[t]
	\centering
\includegraphics[width=1\textwidth]{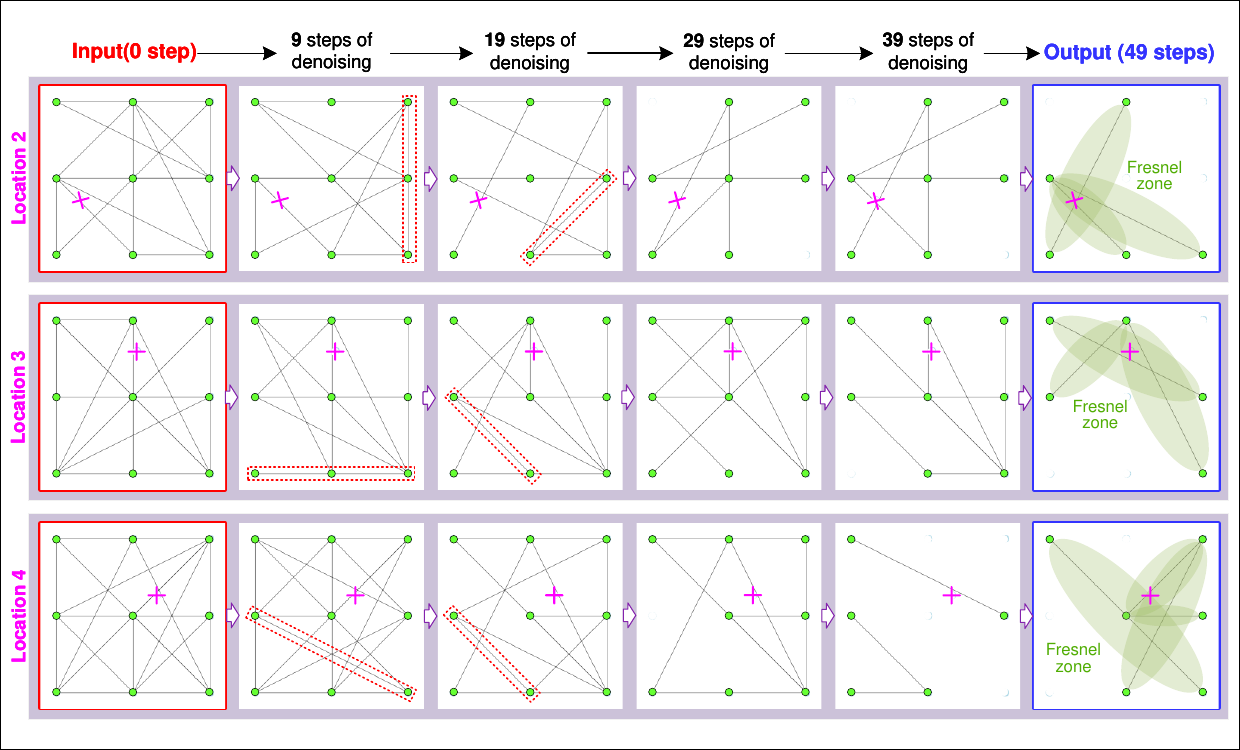}%
	\caption{The denoising process of the trained network. Green dots represent nodes, purple cross indicates the target location, black lines represent activated links, and red dashed lines mark links that contribute little to sensing. The shaded area is the Fresnel zone formed by the activated link. Here, a total of 50 denoising steps are executed to obtain the final graph, with the results at steps 0, 10, 20, 30, 40, and 50 being displayed. }
	\label{GRAG}
\end{figure*}

Figure~\ref{GRAG} presents the graph generation process of the trained denoising network based on the input target locations. As shown, when the number of denoising steps is less than 20, there are many links in the graph, as marked by the red dashed lines, which may be less useful for sensing as they are far from the target. However, as the denoising progresses, the network can continuously optimize the input noisy graph by appropriately adding or removing nodes and links, and the number of ineffective links is significantly reduced after completing 30 steps of denoising. Finally, once 50 steps of denoising are completed, a reasonable graph can be obtained, as shown in blue boxes. The links depicted by the graph are distributed around the target, forming a Fresnel zone that covers the target, thereby enabling the ISAC system to perform the sensing accurately and resource-efficiently. Furthermore, we can see that the graphs generated vary significantly based on different input target locations. This indicates that the trained network can adjust its operations on nodes and links based on the input, thereby generating the desired graphs under various conditions. The results evidently demonstrate that our framework has a high degree of generalizability, allowing it to be applied in various applications.

\section{Future Directions}
\subsection{Graph Rules Design}
In next-generation wireless networks, the irregularity and variability in communication device and links add complexity to the graph generation, necessitating more dynamic and adaptable graph design rules to seamlessly adapt to changing network environments. These rules need to capture the complex relationships and dependencies between network entities more effectively. Therefore, future research could focus on discovering and fine-tuning these rules by analyzing historical network data and ongoing network operations. These refined rules can lead to the development of self-optimizing networks that can adjust their operational parameters in real-time for optimal performance.

\subsection{Scalability and Interpretability}
As wireless networks grow in size and complexity, generating graphs with different scales that accurately reflect the topology and dynamic features of these networks is computationally intensive. Additionally, ensuring that these graphs are interpretable is crucial for their practical application. Therefore, it is essential to create models capable of managing the varying amounts of data produced by these networks. This includes enhancing the capability of GAI models and using the distributed computing resources. While for the interpretability, model simplification, feature importance analysis, and data visualization are key strategies.


\subsection{Mulit-modal Graph Generation}
As datasets grow larger and more complex, including various modalities, representing them through graphs becomes a significant challenge. To address this, future work should focus on improving GAI models to better synthesize and integrate different data types, such as visual, textual, and numerical, into clear and understandable graphs. This involves enhancing GAI models by optimizing their architectures and boosting training stability for multi-modal data. Such enhancements will not only allow these models to handle larger datasets but also improve the accuracy of their representations. These developments are crucial for generating multi-modal graphs that meet the needs of diverse fields such as healthcare and autonomous driving, where it is essential to understand complex data relationships.

\section{Conclusions}
This article investigates the crucial role of graphs in wireless network optimization and proposes GAI based graph generation for wireless networks. Specifically, we first explore the applications of graphs in network routing, resource optimization, and other fields. Then, we propose a framework comprising a conditional diffusion model and an evaluation network. The diffusion model is used to generate desired graphs based on given conditions, while the evaluation network assesses the generated results and provides feedback for optimizing the parameters of the diffusion model. Here, the conditions and evaluation network are defined by users according to specific network problems. Once trained, the proposed framework can generate appropriate graphs in response to changing input conditions, thereby addressing relevant issues in wireless networks. Finally, the effectiveness of the framework is verified using the link selection in the ISAC system as an example, providing guidance for GAI-based graph generation aimed at wireless networks.
\bibliographystyle{IEEEtran}
\bibliography{Ref.bib} 

\begin{thebibliography}{10}
\providecommand{\url}[1]{#1}
\csname url@samestyle\endcsname
\providecommand{\newblock}{\relax}
\providecommand{\bibinfo}[2]{#2}
\providecommand{\BIBentrySTDinterwordspacing}{\spaceskip=0pt\relax}
\providecommand{\BIBentryALTinterwordstretchfactor}{4}
\providecommand{\BIBentryALTinterwordspacing}{\spaceskip=\fontdimen2\font plus
\BIBentryALTinterwordstretchfactor\fontdimen3\font minus
  \fontdimen4\font\relax}
\providecommand{\BIBforeignlanguage}[2]{{%
\expandafter\ifx\csname l@#1\endcsname\relax
\typeout{** WARNING: IEEEtran.bst: No hyphenation pattern has been}%
\typeout{** loaded for the language `#1'. Using the pattern for}%
\typeout{** the default language instead.}%
\else
\language=\csname l@#1\endcsname
\fi
#2}}
\providecommand{\BIBdecl}{\relax}
\BIBdecl

\bibitem{guo2022systematic}
X.~Guo and L.~Zhao, ``A systematic survey on deep generative models for graph
  generation,'' \emph{IEEE Transactions on Pattern Analysis and Machine
  Intelligence}, vol.~45, no.~5, pp. 5370--5390, 2022.

\bibitem{shen2022graph}
Y.~Shen, J.~Zhang, S.~Song, and K.~B. Letaief, ``Graph neural networks for
  wireless communications: From theory to practice,'' \emph{IEEE Transactions
  on Wireless Communications}, 2022.

\bibitem{he2021overview}
S.~He, S.~Xiong, Y.~Ou, J.~Zhang, J.~Wang, Y.~Huang, and Y.~Zhang, ``An
  overview on the application of graph neural networks in wireless networks,''
  \emph{IEEE Open Journal of the Communications Society}, vol.~2, pp.
  2547--2565, 2021.

\bibitem{shi2020graphaf}
C.~Shi, M.~Xu, Z.~Zhu, W.~Zhang, M.~Zhang, and J.~Tang, ``Graphaf: a flow-based
  autoregressive model for molecular graph generation,'' \emph{arXiv preprint
  arXiv:2001.09382}, 2020.

\bibitem{godinez2022design}
W.~J. Godinez, E.~J. Ma, A.~T. Chao, L.~Pei, P.~Skewes-Cox, S.~M. Canham, J.~L.
  Jenkins, J.~M. Young, E.~J. Martin, and W.~A. Guiguemde, ``Design of potent
  antimalarials with generative chemistry,'' \emph{Nature Machine
  Intelligence}, vol.~4, no.~2, pp. 180--186, 2022.

\bibitem{wu2024protein}
K.~E. Wu, K.~K. Yang, R.~van~den Berg, S.~Alamdari, J.~Y. Zou, A.~X. Lu, and
  A.~P. Amini, ``Protein structure generation via folding diffusion,''
  \emph{Nature Communications}, vol.~15, no.~1, p. 1059, 2024.

\bibitem{swaminathan2021graphnet}
A.~Swaminathan, M.~Chaba, D.~K. Sharma, and U.~Ghosh, ``Graphnet: Graph neural
  networks for routing optimization in software defined networks,''
  \emph{Computer Communications}, vol. 178, pp. 169--182, 2021.

\bibitem{shen2020graph}
Y.~Shen, Y.~Shi, J.~Zhang, and K.~B. Letaief, ``Graph neural networks for
  scalable radio resource management: Architecture design and theoretical
  analysis,'' \emph{IEEE Journal on Selected Areas in Communications}, vol.~39,
  no.~1, pp. 101--115, 2020.

\bibitem{you2018graphrnn}
J.~You, R.~Ying, X.~Ren, W.~Hamilton, and J.~Leskovec, ``Graphrnn: Generating
  realistic graphs with deep auto-regressive models,'' in \emph{International
  conference on machine learning}.\hskip 1em plus 0.5em minus 0.4em\relax PMLR,
  2018, pp. 5708--5717.

\bibitem{simonovsky2018graphvae}
M.~Simonovsky and N.~Komodakis, ``Graphvae: Towards generation of small graphs
  using variational autoencoders,'' in \emph{Artificial Neural Networks and
  Machine Learning--ICANN 2018: 27th International Conference on Artificial
  Neural Networks, Rhodes, Greece, October 4-7, 2018, Proceedings, Part I
  27}.\hskip 1em plus 0.5em minus 0.4em\relax Springer, 2018, pp. 412--422.

\bibitem{madhawa2019graphnvp}
K.~Madhawa, K.~Ishiguro, K.~Nakago, and M.~Abe, ``Graphnvp: An invertible flow
  model for generating molecular graphs,'' \emph{arXiv preprint
  arXiv:1905.11600}, 2019.

\bibitem{de2019molgan}
N.~De~Cao and T.~Kipf, ``Molgan: An implicit generative model for small
  molecular graphs. arxiv 2018,'' \emph{arXiv preprint arXiv:1805.11973}, 2019.

\bibitem{zhang2023survey}
M.~Zhang, M.~Qamar, T.~Kang, Y.~Jung, C.~Zhang, S.-H. Bae, and C.~Zhang, ``A
  survey on graph diffusion models: Generative ai in science for molecule,
  protein and material,'' \emph{arXiv preprint arXiv:2304.01565}, 2023.

\bibitem{liu2024graph}
Y.~Liu, C.~Du, T.~Pang, C.~Li, W.~Chen, and M.~Lin, ``Graph diffusion policy
  optimization,'' \emph{arXiv preprint arXiv:2402.16302}, 2024.

\bibitem{wang2024unified}
J.~Wang, H.~Du, D.~Niyato, J.~Kang, Z.~Xiong, D.~Rajan, S.~Mao, and X.~Shen,
  ``A unified framework for guiding generative ai with wireless perception in
  resource constrained mobile edge networks,'' \emph{IEEE Transactions on
  Mobile Computing}, 2024.

\end{thebibliography}

\end{document}